\documentclass[prd,showkeys,floatfix,twocolumn,amsmath,amssymb,floatfix]{revtex4}
\usepackage{graphicx,color,dcolumn,booktabs,bm}
\usepackage{longtable,lscape}
\usepackage{amssymb}
\usepackage{amsmath}
\usepackage{indentfirst}
\usepackage{epsfig}
\usepackage{feynmf}
\usepackage{subfigure}
\usepackage{epstopdf}
\usepackage{slashed}
\usepackage{cases}
\usepackage{multirow}
\usepackage[colorlinks,citecolor=blue,anchorcolor=red,menucolor=red,linkcolor=red,filecolor=red,runcolor=red,urlcolor=blue,frenchlinks=red]{hyperref}

\allowdisplaybreaks[3]

\begin{document}

\title{Topped baryons from QCD sum rules}

\author{Shu-Wei Zhang}
\author{Wei-Han Tan}
\author{Xuan Luo}
\author{Hua-Xing Chen}
\email{hxchen@seu.edu.cn}

\affiliation{School of Physics, Southeast University, Nanjing 210094, China}

\begin{abstract}
The recent CMS observation of a near-threshold enhancement in top quark pair production~\cite{CMS:2025kzt} provides the first experimental indication of a short-lived pseudoscalar $t\bar{t}$ bound state, commonly referred to as \textit{toponium}. While most existing studies focus on toponium using perturbative QCD near threshold, baryonic configurations containing a single top quark—\textit{singly topped baryons}—offer a complementary nonperturbative perspective. Notably, singly topped baryons are expected to exhibit a longer lifetime and a narrower decay width than toponium, since only one top quark decays rather than two. The heavy quark effective theory provides a natural and powerful framework for analyzing such systems, allowing the separation of heavy and light quark dynamics. In this work we employ heavy quark effective theory to investigate the internal structure of ground-state singly topped baryons. We construct their interpolating currents and analyze them using QCD sum rules. The resulting masses of the ground-state singly topped baryons are found to lie around 174~GeV, approximately $1.1$–$1.5$~GeV above the pole mass of the top quark. In parallel, topped mesons are systematically examined in Ref.~\cite{Zhang:2025fdp} using the same theoretical approach.
\end{abstract}


\maketitle

\pagenumbering{arabic}

\section{Introduction}

The top quark, with its exceptionally large mass of approximately 173~GeV, stands apart from all other quarks in the Standard Model. Unlike lighter quarks, the top quark decays before it can hadronize due to its ultrashort lifetime of about $5 \times 10^{-25}$ seconds, presenting both a challenge and an opportunity for hadronic physics. Traditionally, most theoretical attention has focused on the \textit{toponium} system—a hypothetical $t\bar{t}$ bound state. Although such a state cannot form a stable spectrum due to the top quark's rapid decay, near-threshold $t\bar{t}$ production remains governed by nonrelativistic QCD dynamics. Extensive calculations based on perturbative QCD and potential models predict distinct enhancements near the production threshold~\cite{Fadin:1987wz,Kuhn:1987ty,Barger:1987xg,Fadin:1990wx,Strassler:1990nw,Sumino:1997ve,Hoang:2000yr,Penin:2005eu,Kiyo:2008bv,Hagiwara:2008df,Sumino:2010bv,Beneke:2015kwa,Fuks:2021xje,Akbar:2024brg,Wang:2024hzd,Jiang:2024fyw,Fuks:2024yjj}. Notably, the CMS Collaboration has recently reported an excess in the $t\bar{t}$ invariant mass spectrum near threshold~\cite{CMS:2025kzt}, with properties consistent with the predicted pseudoscalar toponium state; see also the earlier ATLAS and CMS results~\cite{ATLAS:2023fsd,CMS:2024pts}. The observed excess shows significantly better consistency with QCD toponium predictions than with alternative interpretations, potentially representing the first experimental evidence of toponium formation and marking a pivotal advancement in top-quark physics~\cite{Jafari:2025rmm,Aguilar-Saavedra:2024mnm,Llanes-Estrada:2024phk,Nason:2025hix,Ellis:2025nkm,Fu:2025yft,Fu:2025zxb,Bai:2025buy}.

Complementary to the toponium system is a less-explored class of baryonic configurations known as \textit{singly topped baryons}, consisting of one heavy top quark and two light quarks (up, down, or strange). Although these systems are not expected to form bound states in practice due to the ultrashort lifetime of the top quark, they nevertheless offer a valuable and distinct theoretical perspective. One notable advantage of singly topped baryons over toponium lies in their decay dynamics and experimental signatures. In toponium, the presence of both a top and an antitop quark enables annihilation channels, which significantly broaden its decay width through strong and electroweak interactions. In contrast, singly topped baryons are free from such annihilation effects. Given that only one top quark decays, these baryons are theoretically expected to exhibit a lifetime roughly twice that of toponium, resulting in a correspondingly narrower decay width~\cite{Chen:2021erj,Maltoni:2024csn}.

It is also instructive to compare these expectations with known broad hadronic resonances such as the light scalar mesons $\sigma(500)$ and $\kappa(800)$, which have estimated decay widths of $400$--$700\,\mathrm{MeV}$ and $500$--$800\,\mathrm{MeV}$, respectively~\cite{pdg}. These large widths have historically complicated both identification and precise line-shape analysis. By contrast, the top quark has a well-measured decay width of $\Gamma_t \approx 1.41\,\mathrm{GeV}$~\cite{pdg}, and a singly topped baryon would inherit this width without additional broadening from $t\bar{t}$ annihilation. From both theoretical and phenomenological perspectives, singly topped baryons therefore represent a compelling opportunity to probe top-quark dynamics beyond the conventional $t\bar{t}$ framework.

Importantly, singly topped baryons are amenable to treatment within the framework of heavy quark effective theory (HQET)~\cite{Neubert:1993mb, Manohar:2000dt}. In this approach, the top quark is approximated as a static color source, enabling a systematic classification of the light diquark subsystem according to its color, flavor, spin, and orbital quantum numbers. This simplification facilitates the construction of interpolating currents suitable for use in QCD sum rule analyses and lattice QCD simulations~\cite{Dai:1993kt, Dai:1996yw, Liu:2007fg,Yang:2025hzc}. In this paper we construct and analyze the interpolating currents for ground-state singly topped baryons within the heavy quark effective theory framework. Based on a detailed classification of the light-quark degrees of freedom, we derive appropriate current structures and perform QCD sum rule analyses to investigate their properties. The resulting masses are found to lie around 174~GeV, approximately $1.1$–$1.5$~GeV above the pole mass of the top quark.

\begin{figure*}[htb]
\begin{center}
\scalebox{0.95}{\includegraphics{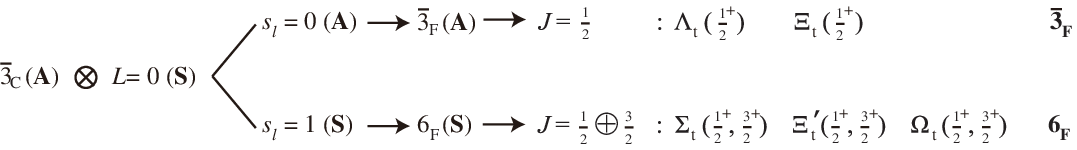}}
\end{center}
\caption{Categorization of singly topped baryons.
\label{fig:topbaryon}}
\end{figure*}

This paper is organized as follows. In Sec.~\ref{sec:current} we systematically construct the interpolating currents for ground-state singly topped baryons. These currents are then employed in Sec.~\ref{sec:leading} to perform QCD sum rule analyses at the leading order. Next, in Sec.~\ref{sec:nexttoleading} we incorporate ${\mathcal O}(1/m_Q)$ corrections to examine next-to-leading-order effects. A summary and discussion of our findings are presented in Sec.~\ref{sec:summary}.

\section{Interpolating currents for ground-state topped baryons}
\label{sec:current}

In this section we briefly introduce the notations and conventions used throughout this work. A singly topped baryon is described as a three-quark system composed of two light quarks—up, down, or strange—and one heavy top quark. To simplify the presentation, we shall refer to a \textit{singly topped baryon} simply as a \textit{topped baryon} in the remainder of this paper.

To properly understand its internal structure, it is essential to analyze the quantum numbers and symmetries associated with the two light quarks, including their color, flavor, spin, and orbital degrees of freedom. These components are constrained by the Pauli exclusion principle, which requires the total wave function of the two light quarks to be antisymmetric under exchange. The structural features are summarized as follows:
\begin{itemize}
    
\item The two light quarks must form an antisymmetric color configuration, constituting a color antitriplet ($\bm{\bar{3}}_C$).

\item The flavor wave function can be either antisymmetric, corresponding to the $SU(3)$ flavor antitriplet ($\bm{\bar{3}}_F$), or symmetric, corresponding to the $SU(3)$ flavor sextet ($\bm{6}_F$).

\item The spin configuration of the two light quarks can be antisymmetric ($s_l = 0$) or symmetric ($s_l = 1$), where $s_l$ denotes the total spin of the light diquark system.

\item The orbital configuration of the two light quarks is symmetric in ground-state topped baryons.

\end{itemize}
As illustrated in Fig.~\ref{fig:topbaryon}, the ground-state topped baryons can be systematically categorized into two distinct HQET multiplets. One belongs to the $SU(3)$ flavor $\bm{\bar{3}}_F$ representation, while the other belongs to the $SU(3)$ flavor $\bm{6}_F$ representation. For each HQET multiplet, the total spin of the baryon is obtained by coupling $s_l$ with the top quark spin $s_t = 1/2$, resulting in physical states with $J = |s_l \pm 1/2|$.

The interpolating currents associated with ground-state bottom baryons have been systematically constructed in Ref.~\cite{Liu:2007fg}. In the present study, we replace the bottom quark with the top quark and construct the corresponding currents for ground-state topped baryons. In general, the interpolating current for a topped baryon can be expressed as a combination of a light diquark field and a heavy top quark field:
\begin{eqnarray}
J(x) \sim \epsilon_{abc} \left[ q^{aT}(x)\, C\, \Gamma_1\, q^b(x) \right] \Gamma_2\, h_v^c(x)\, ,
\label{eq:baryonfield}
\end{eqnarray}
where $a$, $b$, and $c$ are color indices, and $\epsilon_{abc}$ is the totally antisymmetric tensor. The superscript $T$ denotes transpose with respect to Dirac indices only, and $C$ is the charge-conjugation matrix. The field $q(x)$ represents a light quark at position $x$, which may be $u(x)$, $d(x)$, or $s(x)$. The field $h_v(x)$ denotes the top quark in the HQET framework, and we have applied the Fierz transformation to move it to the rightmost position. Additionally, we adopt the projections
\[
\gamma_t^\mu = \gamma^\mu - v\!\!\!\slash\, v^\mu\,, \qquad g_t^{\alpha_1\alpha_2} = g^{\alpha_1\alpha_2} - v^{\alpha_1}v^{\alpha_2}\,,
\]
where $v^\mu$ is the four-velocity of the heavy top quark.

There are two ``good'' $S$-wave diquark fields. The first is
\begin{eqnarray}
\epsilon_{abc}\, q^{aT}(x)\, C\, \gamma_5\, q^b(x) \qquad [^1S_0]\,,
\end{eqnarray}
which has quantum numbers $j_l^{P_l} = 0^+$. Its orbital wave function is symmetric. The spin configuration is $s_l = 0$ and hence antisymmetric, while the color structure is $\bm{\bar{3}}_C$ and also antisymmetric. Consequently, the Pauli exclusion principle requires the flavor structure to be antisymmetric as well, corresponding to $\bm{\bar{3}}_F$.

The second $S$-wave diquark field is
\begin{eqnarray}
\epsilon_{abc}\, q^{aT}(x)\, C\, \gamma_\mu\, q^b(x) \qquad [^3S_1]\,,
\end{eqnarray}
which carries quantum numbers $j_l^{P_l} = 1^+$. It has $l_\rho = 0$ (symmetric, $\mathbf{S}$), $s_l = 1$ (symmetric, $\mathbf{S}$), color $\bm{\bar{3}}_C$ (antisymmetric, $\mathbf{A}$), and flavor $\bm{6}_F$ (symmetric, $\mathbf{S}$).

These diquark fields can be used to construct interpolating currents for ground-state topped baryons, whose explicit forms are given below:
\begin{itemize}

\item When the diquark has $s_l = 0$ ($\mathbf{A}$), its flavor representation is $\bm{\bar{3}}_F$ ($\mathbf{A}$). The total angular momentum of the baryon is $J = s_l \otimes s_t = 1/2$, corresponding to an HQET singlet with $J^P = 1/2^+$:
\begin{eqnarray}
J_{\bm{\bar{3}}F}(x) &=& \epsilon_{abc}\, \left[q^{aT}(x) C \gamma_5 q^b(x)\right]\, h_v^c(x)\, .
\label{def:J3}
\end{eqnarray}

\item When the diquark has $s_l = 1$ ($\mathbf{S}$), its flavor representation is $\bm{6}_F$ ($\mathbf{S}$). The total angular momentum becomes $J = s_l \otimes s_t = 1/2 \oplus 3/2$, yielding an HQET doublet with $J^P = (1/2^+,\, 3/2^+)$:
\begin{eqnarray}
J_{\bm{6}F}(x) &=& \epsilon_{abc}\, \left[q^{aT}(x) C \gamma_\mu q^b(x)\right]\, \gamma_t^\mu \gamma_5 h_v^c(x)\, ,
\label{def:J6one}
\\ \nonumber
J^\mu_{\bm{6}F}(x) &=& \epsilon_{abc}\, \left[q^{aT}(x) C \gamma_\nu q^b(x)\right] \left(g_t^{\mu\nu} - \frac{\gamma_t^\mu\gamma_t^\nu}{3} \right) h_v^c(x)\, .
\\ \label{def:J6three}
\end{eqnarray}

\end{itemize}
Identical QCD sum rules can be obtained using either $J_{\bm{6}F}(x)$ or $J^\mu_{\bm{6}F}(x)$, as they belong to the same HQET doublet~\cite{Dai:1993kt,Dai:1996yw,Liu:2007fg,Dai:1996qx,Dai:2003yg}. Therefore, it is sufficient to use one of them in the QCD sum rule analysis. In the following sections, we adopt $J_{\bm{\bar{3}}F}(x)$ and $J_{\bm{6}F}(x)$ to study baryon multiplets in the $SU(3)$ flavor representations $\bm{\bar{3}}_F$ and $\bm{6}_F$, respectively.

Before proceeding, we explicitly specify the quark content of the interpolating currents. For example, $J_{\Xi^\prime_t}(x)$ denotes the current $J_{\bm{6}F}(x)$ with the flavor structure $qst$ ($q=u/d$):
\begin{eqnarray}
J_{\Xi^\prime_t}(x) &=& \epsilon_{abc}\, \left[q^{aT}(x) C \gamma_\mu s^b(x)\right]\, \gamma_t^\mu \gamma_5 h_v^c(x)\, .
\label{eq:example}
\end{eqnarray}
In the subsequent sections, we use this current as a representative example in the QCD sum rule analysis of the topped baryon $\Xi^\prime_t$ with $J^P = 1/2^+$.

\section{Results at the leading order}
\label{sec:leading}

In this section we apply the QCD sum rule method to investigate the interpolating currents for ground-state topped baryons, as classified in the previous section. As a representative example, we consider the current \( J_{\Xi^\prime_t}(x) \), whose coupling to the physical baryon \( \Xi^\prime_t \) is defined as
\begin{equation}
\langle 0 | J_{\Xi^\prime_t}(x) | \Xi^\prime_t \rangle = f \times u(x) \, ,
\end{equation}
where \( f \) is the decay constant and \( u(x) \) is the Dirac spinor. This leads to the construction of the two-point correlation function:
\begin{equation}
\Pi(\omega) = i \int d^4 x\, e^{i k \cdot x} \langle 0 |
T[J_{\Xi^\prime_t}(x) \bar J_{\Xi^\prime_t}(0)] | 0 \rangle \, ,
\label{eq:pi}
\end{equation}
where \( \omega = v \cdot k \) denotes the off-shell energy.

At the hadronic level, Eq.~(\ref{eq:pi}) can be expressed as
\begin{equation}
\Pi(\omega) = \frac{f^{2}}{\overline{\Lambda} - \omega} + \text{higher resonances} \, ,
\label{eq:pole}
\end{equation}
where
\begin{equation}
\overline{\Lambda} \equiv \lim_{m_t \rightarrow \infty} (m_{\Xi^\prime_t} - m_t) \, ,
\label{eq:leading}
\end{equation}
denotes the residual mass of the \( \Xi^\prime_t \) baryon in the heavy quark limit, with \( m_{\Xi^\prime_t} \) representing its physical mass.

On the QCD side, Eq.~(\ref{eq:pi}) is evaluated using the operator product expansion (OPE). Substituting Eq.~(\ref{eq:example}) and applying a Borel transformation, we obtain
\begin{align}
\Pi(\omega_c, T) &= f(\omega_c, T)^2 \cdot e^{- \overline{\Lambda}(\omega_c, T) / T} \nonumber\\
&= \int_{s_<}^{\omega_c} \left[\frac{3\omega^{5}}{20\pi^4}
+\frac{(3m_{q} m_{s}-3m_{q}^{2}-3m_{s}^2)
\omega^{3}}{4\pi^{4}} \right. \nonumber\\
&\quad \left. -\frac{\langle g^2 GG \rangle \omega}{128\pi^4}
-\frac{6m_{q}\langle\bar{s}s\rangle +6m_{s}\langle\bar{q}q\rangle}{4\pi^{2}}\omega \right. \nonumber\\
&\quad \left. +\frac{3m_{q}\langle\bar{q}q\rangle +3m_{s}\langle\bar{s}s\rangle}{4\pi^{2}}\omega \right]
e^{-\omega/T} \, d\omega \nonumber\\
&\quad +\frac{\langle\bar{q}q\rangle\langle\bar{s}s\rangle}{2}
- \frac{3m_{q} \langle g_s \bar{s} \sigma G s \rangle + 3m_{s} \langle g_s \bar{q} \sigma G q \rangle}{32\pi^2} \nonumber\\
&\quad + \frac{5m_{q} \langle g_s \bar{q} \sigma G q \rangle + 5m_{s} \langle g_s \bar{s} \sigma G s \rangle}{128\pi^2} \nonumber\\
&\quad + \frac{
\langle\bar{s}s\rangle\langle g_s\bar{q}\sigma G q\rangle +
\langle\bar{q}q\rangle\langle g_s\bar{s}\sigma G s\rangle }{32T^2} \, .
\label{eq:ope}
\end{align}
In the above expression, \( s_< = m_s + m_q \) is the physical threshold, \( \omega_c \) is the continuum threshold value, and \( T \) is the Borel mass. The introduction of \( \omega_c \) reflects the assumption of quark–hadron duality: contributions from higher resonances and continuum states above \( \omega_c \) are modeled by the perturbative part of the OPE. In practice, \( \omega_c \) is chosen to be moderately above the residual mass of the target baryon to ensure pole dominance and good convergence of the OPE series.

\begin{figure*}[htbp]
\centering
\subfigure[]{\scalebox{0.42}{\includegraphics{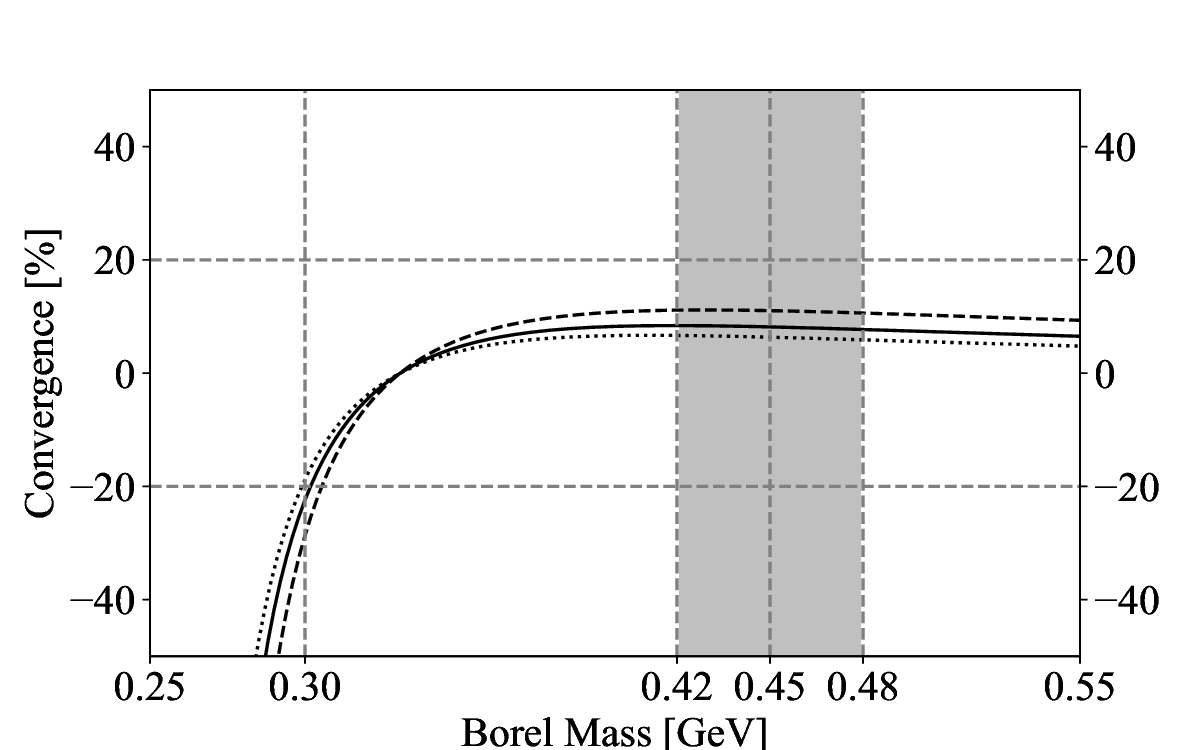}}}
\hspace{0.5cm}
\subfigure[]{\scalebox{0.42}{\includegraphics{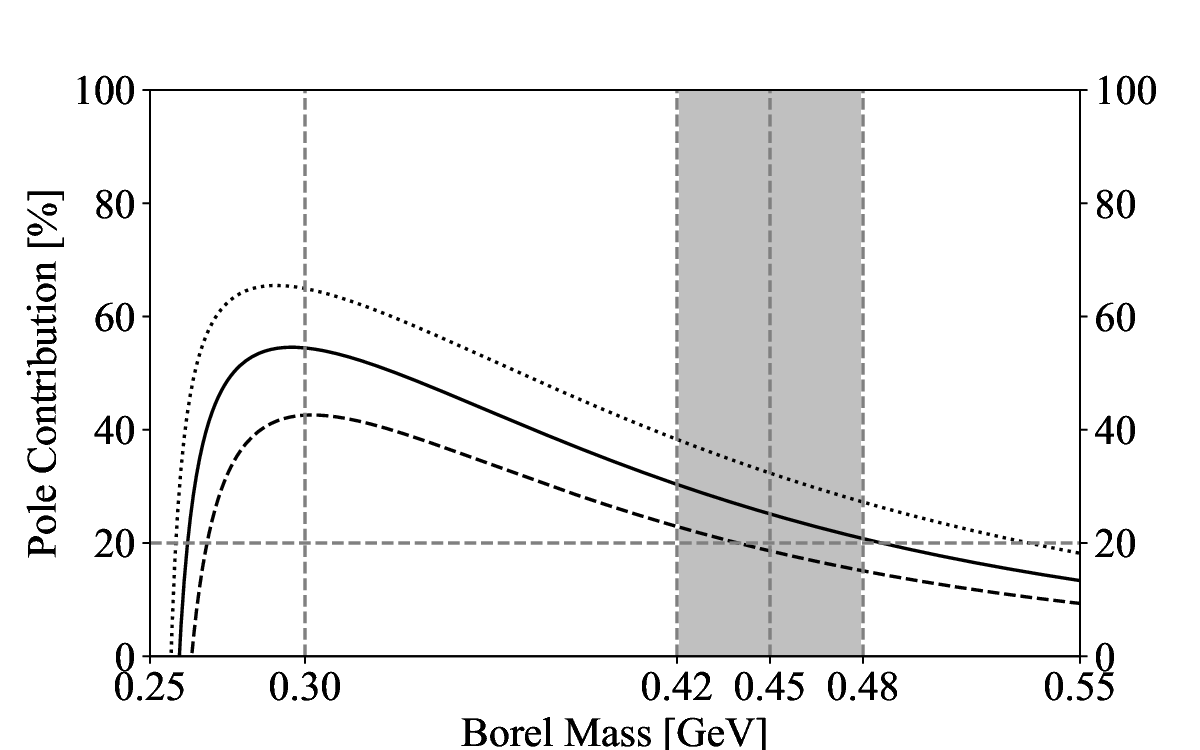}}}
\caption{Variations of (a) CVG and (b) PC, as defined in Eqs.~(\ref{eq_convergence}) and (\ref{eq_pole}), with respect to the Borel mass $T$, using the interpolating current $J_{\Xi^\prime_t}(x)$. The dashed, solid, and dotted curves correspond to $\omega_c = 1.65$, $1.85$, and $2.05~\mathrm{GeV}$, respectively.}
\label{fig:pole}
\end{figure*}

\begin{figure*}[htbp]
\centering
\subfigure[]{\scalebox{0.42}{\includegraphics{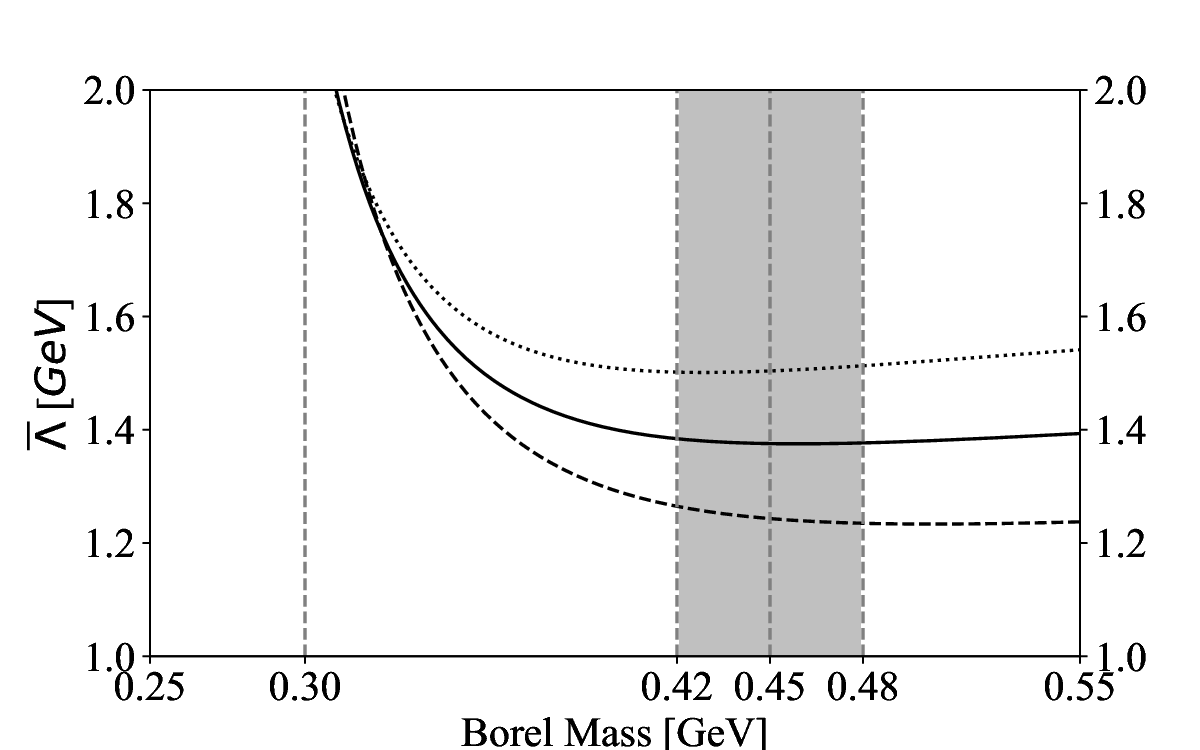}}}
\hspace{0.5cm}
\subfigure[]{\scalebox{0.42}{\includegraphics{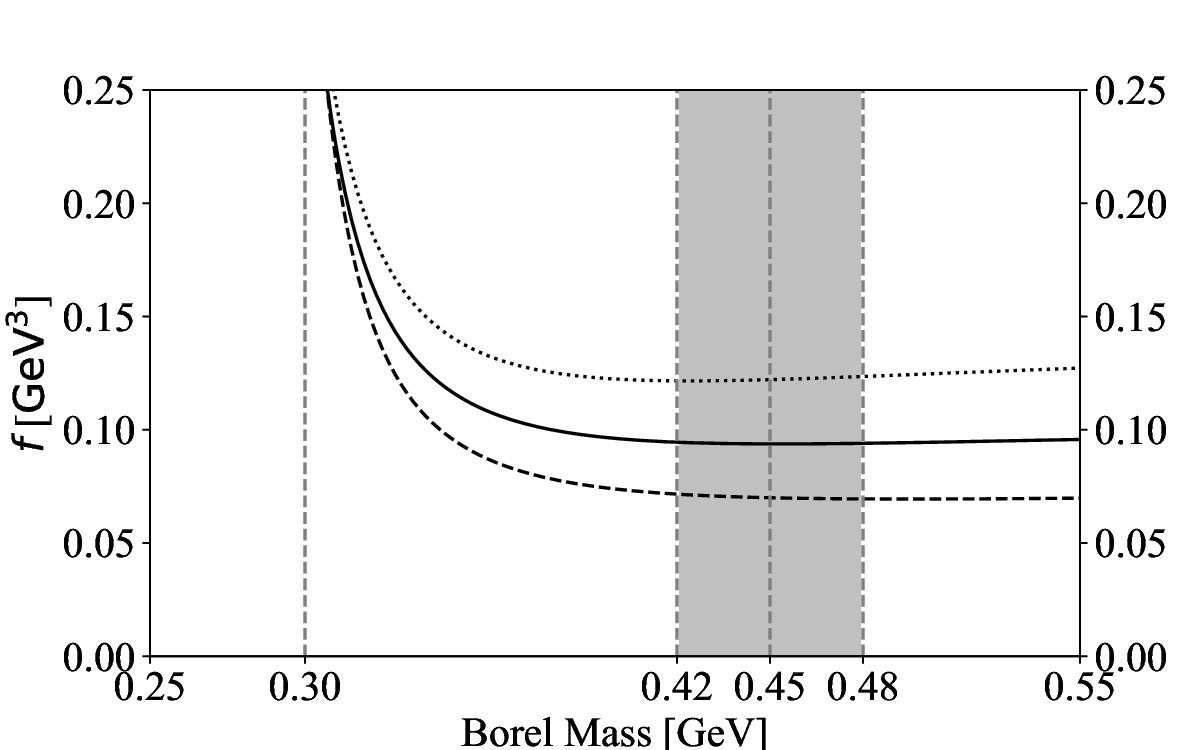}}}
\caption{Dependence of (a) the residual mass $\overline{\Lambda}$ and (b) the decay constant $f$ on the Borel mass $T$, using the interpolating current $J_{\Xi^\prime_t}(x)$. The dashed, solid, and dotted curves correspond to $\omega_c = 1.65$, $1.85$, and $2.05~\mathrm{GeV}$, respectively.}
\label{fig:leading}
\end{figure*}

The physical observables can be extracted by differentiating Eq.~(\ref{eq:ope}) with respect to $-1/T$, yielding
\begin{eqnarray}
\overline{\Lambda}(\omega_c, T)
&=& \frac{1}{\Pi(\omega_c, T)} \cdot \frac{\partial \Pi(\omega_c, T)}{\partial(-1/T)} \, ,
\label{eq:mass}
\\
f(\omega_c, T)^2
&=& \Pi(\omega_c, T) \cdot e^{\overline{\Lambda}(\omega_c, T) / T} \, .
\label{eq:coupling}
\end{eqnarray}
There are two free parameters in these expressions: the threshold value $\omega_c$ and the Borel mass $T$. To determine their reliable working windows, we impose three standard criteria. Throughout this analysis, we adopt the following values for the QCD input parameters at the renormalization scale $\mu = 2~\mathrm{GeV}$~\cite{pdg,Yang:1993bp,Narison:2011xe,Narison:2018dcr,Gimenez:2005nt,Jamin:2002ev,Ioffe:2002be,Ovchinnikov:1988gk,Colangelo:2000dp}:
\begin{eqnarray}
\nonumber \langle \bar{q} q \rangle &=& -(0.240~\mathrm{GeV})^3 \, , 
\\ \nonumber \langle \bar{s} s \rangle &=& (0.8 \pm 0.1) \times \langle \bar{q} q \rangle \, , 
\\ \nonumber \langle g_s \bar{q} \sigma G q \rangle &=& - M_0^2 \times \langle \bar{q} q \rangle \, , 
\\ \langle g_s \bar{s} \sigma G s \rangle &=& - M_0^2 \times \langle \bar{s} s \rangle \, ,
\label{eq:condensate}
\\ \nonumber M_0^2 &=& (0.8 \pm 0.2)~\mathrm{GeV}^2 \, , 
\\ \nonumber \langle g_s^2 G^2 \rangle &=& (0.48 \pm 0.14) ~\mathrm{GeV}^4 \, , 
\\ \nonumber m_s &=& 93.5 \pm 0.8~\mathrm{MeV} \, ,
\\ \nonumber m_q &\approx& m_u \approx m_d \approx 0~\mathrm{MeV} \, .
\end{eqnarray}

The first criterion requires that the high-dimensional power corrections contribute less than 20\% of the total correlator:
\begin{equation}
\label{eq_convergence}
\mathrm{CVG} \equiv \left| \frac{ \Pi^{\text{high-order}}(\omega_c, T) }{ \Pi(\omega_c, T) } \right| \leq 20\% \, .
\end{equation}
As shown in Fig.~\ref{fig:pole}(a), this sets the lower bound of the Borel mass as $T_{\min} = 0.30~\mathrm{GeV}$.

The second criterion requires the pole contribution (PC) to be no less than 20\%:
\begin{equation}
\label{eq_pole}
\mathrm{PC} \equiv \frac{ \Pi(\omega_c, T) }{ \Pi(\infty, T) } \geq 20\% \, .
\end{equation}
As shown in Fig.~\ref{fig:pole}(b), this determines the upper bound $T_{\max} = 0.48~\mathrm{GeV}$ for $\omega_c = 1.85~\mathrm{GeV}$. Together, these two conditions define a preliminary Borel window:
\begin{equation}
0.30~\mathrm{GeV} < T < 0.48~\mathrm{GeV} \, .
\end{equation}

To further refine this range, we analyze the $T$-dependence of the extracted parameters $\overline{\Lambda}$ and $f$, as shown in Fig.~\ref{fig:leading}. The convergence ratio CVG peaks near $T_{\rm peak} = 0.42~\mathrm{GeV}$, above which the extracted results become significantly more stable. Below this point, both $\overline{\Lambda}$ and $f$ exhibit strong sensitivity to $T$, violating the stability requirement. Therefore, the third criterion—Borel stability—restricts the working region to
\begin{equation}
0.42~\mathrm{GeV} < T < 0.48~\mathrm{GeV} \, .
\end{equation}
Within this refined Borel window, the extracted numerical values are
\begin{eqnarray}
\overline{\Lambda} &=& 1.38_{-0.14}^{+0.14}~\mathrm{GeV} \, , \\
f &=& 0.09_{-0.02}^{+0.03}~\mathrm{GeV}^3 \, ,
\end{eqnarray}
where the central values correspond to $T = 0.45~\mathrm{GeV}$ and $\omega_c = 1.85~\mathrm{GeV}$.

\section{Results at the ${\mathcal O}(1/m_Q)$ order}
\label{sec:nexttoleading}

In this section we extend the analysis to include ${\mathcal O}(1/m_Q)$ corrections~\cite{Dai:1996qx,Dai:2003yg}. To this end, we employ the effective Lagrangian of HQET:
\begin{eqnarray}
\mathcal{L}_{\rm eff} = \bar{h}_{v}^a i v \cdot D\, h_{v}^a + \frac{1}{2m_t} \mathcal{K} + \frac{1}{2m_t} \mathcal{S} \, ,
\label{eq:next}
\end{eqnarray}
where $\mathcal{K}$ denotes the nonrelativistic kinetic energy operator,
\begin{eqnarray}
\mathcal{K} = \bar{h}_{v}^a (i D_{t})^{2} h_{v}^a \, ,
\end{eqnarray}
and $\mathcal{S}$ represents the Pauli term describing the chromomagnetic interaction,
\begin{eqnarray}
\mathcal{S} = \frac{g_s}{2} C_{\rm mag} \left( \frac{m_t}{\mu} \right) \bar{h}_{v}^a \sigma_{\mu\nu} G^{\mu\nu} h_{v}^a \, ,
\end{eqnarray}
with the Wilson coefficient
\begin{eqnarray}
C_{\rm mag} \left( \frac{m_t}{\mu} \right) = \left[ \frac{\overline{\alpha}_s(m_t)}{\overline{\alpha}_s(\mu)} \right]^{3/\beta_0} \, ,
\end{eqnarray}
and $\beta_0 = 11 - \frac{2}{3} n_f$. At the renormalization scale $\mu = 2~\mathrm{GeV}$, this coefficient evaluates to $C_{\rm mag}(\mu) \approx 0.6$ for the top quark.

The pole term can then be expanded up to ${\mathcal O}(1/m_Q)$ as
\begin{eqnarray}
\Pi(\omega) &=& \frac{(f + \delta f)^2}{\overline{\Lambda} + \delta m - \omega}
\label{eq:correction}
\\ \nonumber &=& \frac{f^2}{\overline{\Lambda} - \omega} - \frac{\delta m f^2}{(\overline{\Lambda} - \omega)^2} + \frac{2f \delta f}{\overline{\Lambda} - \omega} \, ,
\end{eqnarray}
where $\delta m$ and $\delta f$ denote the ${\mathcal O}(1/m_Q)$ corrections to the baryon mass $m_{\Xi^\prime_t}$ and the decay constant $f$, respectively.

To evaluate $\delta m$, we consider the following three-point correlation functions:
\begin{eqnarray}
\delta_O \Pi(\omega, \omega^\prime)
&=& i^2 \int d^4x\, d^4y\, e^{i k \cdot x - i k^\prime \cdot y}
\label{eq:nextpi}
\\ \nonumber && ~~~~~ \times \langle 0 | T[J_{\Xi^\prime_t}(x)\, O(0)\, \bar{J}_{\Xi^\prime_t}(y)] | 0 \rangle \, ,
\end{eqnarray}
where $O = \mathcal{K}$ or $\mathcal{S}$. Based on the effective Lagrangian in Eq.~(\ref{eq:next}), the above correlators can be expressed at the hadron level as
\begin{eqnarray}
\delta_{\mathcal{K}} \Pi(\omega, \omega^\prime) &=& \frac{f^2 K}{(\overline{\Lambda} - \omega)(\overline{\Lambda} - \omega^\prime)} + \cdots \, ,
\label{eq:K}
\\
\delta_{\mathcal{S}} \Pi(\omega, \omega^\prime) &=& \frac{d_M f^2 \Sigma}{(\overline{\Lambda} - \omega)(\overline{\Lambda} - \omega^\prime)} + \cdots \, ,
\label{eq:S}
\end{eqnarray}
where $K$, $\Sigma$, and $d_M$ are defined as
\begin{eqnarray}
K &\equiv& \langle \Xi^\prime_t | \bar{h}_v^a (i D_\perp)^2 h_v^a | \Xi^\prime_t \rangle \, ,
\\
d_M \Sigma &\equiv& \left\langle \Xi^\prime_t \left| \frac{g_s}{2} \bar{h}_v^a \sigma_{\mu\nu} G^{\mu\nu} h_v^a \right| \Xi^\prime_t \right\rangle \, ,
\\
d_M &\equiv& d_{j, j_l} \, ,
\\
d_{j_l - 1/2, j_l} &=& 2 j_l + 2 \, ,
\\
d_{j_l + 1/2, j_l} &=& -2 j_l \, .
\end{eqnarray}
Setting $\omega = \omega^\prime$ and combining Eqs.~(\ref{eq:correction}), (\ref{eq:K}), and (\ref{eq:S}), we obtain
\begin{eqnarray}
\delta m = -\frac{1}{2 m_t} \left( K + d_M C_{\rm mag} \Sigma \right) \, .
\label{eq:more}
\end{eqnarray}

\begin{figure*}[hbt]
\begin{center}
\subfigure[]{\scalebox{0.42}{\includegraphics{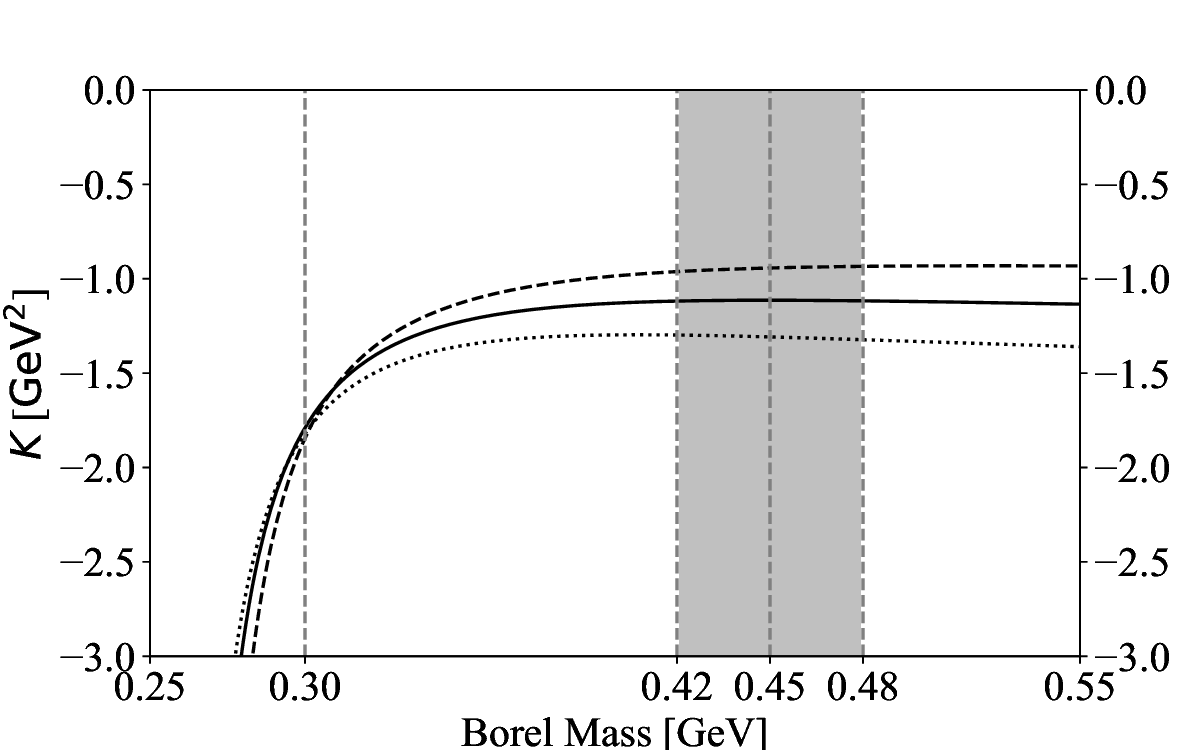}}}
~~~~~
\subfigure[]{\scalebox{0.42}{\includegraphics{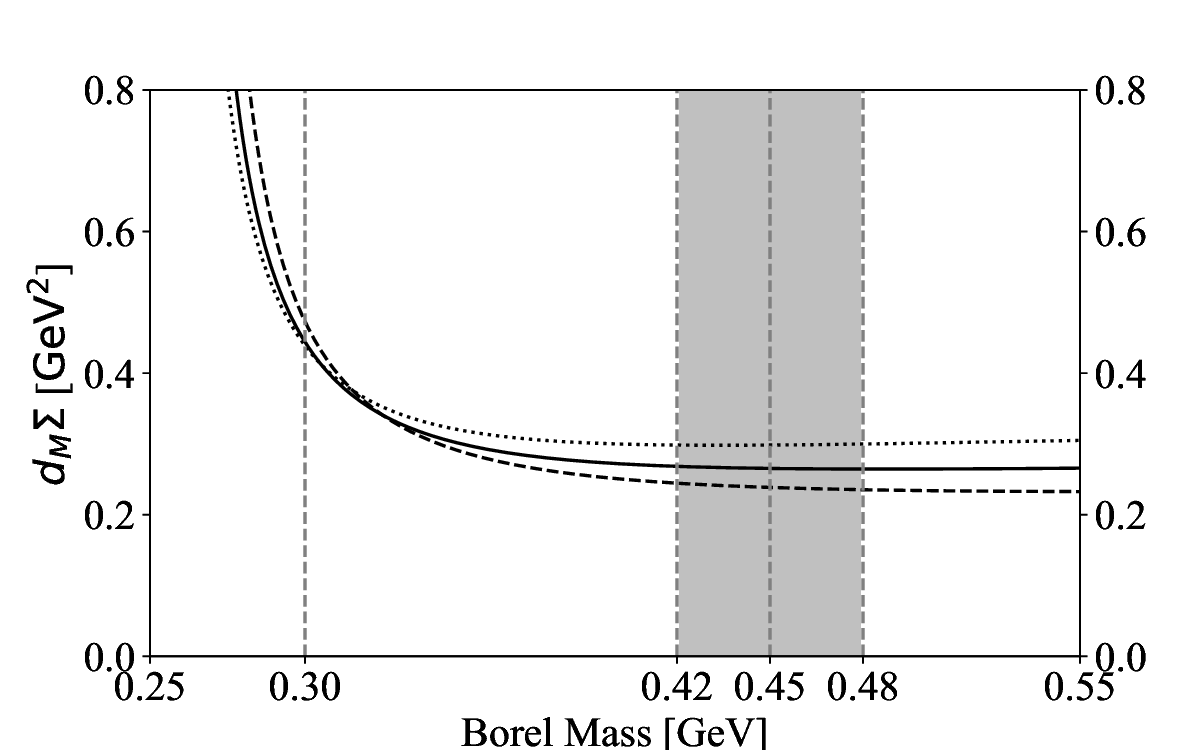}}}
\caption{Variations of (a) $K$ and (b) $d_M\Sigma$ with respect to the Borel mass
$T$, using the interpolating current $J_{\Xi^\prime_t}(x)$. The working region is $0.42~\mathrm{GeV} < T < 0.48~\mathrm{GeV}$. The dashed, solid, and dotted curves correspond to $\omega_c = 1.65$, $1.85$, and $2.05~\mathrm{GeV}$, respectively.}
\label{fig:KandS}
\end{center}
\end{figure*}

The correlation functions in Eq.~(\ref{eq:nextpi}) can also be evaluated at the quark-gluon level using the operator product expansion. By substituting Eq.~(\ref{eq:example}) into Eq.~(\ref{eq:nextpi}) and performing a double Borel transformation in $\omega$ and $\omega^\prime$, we obtain sum rule equations that depend on two Borel parameters $T_1$ and $T_2$. Setting $T_1 = T_2 = T$, we arrive at
\begin{eqnarray}
&&f^2 K\, e^{-\overline{\Lambda} / T}
\nonumber\\&=&-\Bigg\{\int_{s_<}^{\omega_c}
\bigg[\frac{396 \, \omega^7}{7! \, \pi^4}
-\frac{18 \, \omega^5}{5! \, \pi^4}(4m_q^2 + 4m_s^2 - 3m_q m_s)
\nonumber\\&& -\frac{\langle g^2 GG \rangle \, \omega^3}{384 \, \pi^4}
+\frac{ \, \omega^3}{4 \, \pi^2}
\Big(5m_q \langle \bar{q} q \rangle + 5m_s \langle \bar{s} s \rangle
\nonumber\\&& -6m_s \langle \bar{q} q \rangle - 6m_q \langle \bar{s} s \rangle \Big)
+\frac{11 \, \omega}{256 \, \pi^2}
\Big(m_q \langle g_s \bar{q} \sigma G q \rangle
\nonumber\\&& + m_s \langle g_s \bar{s} \sigma G s \rangle \Big)
\bigg] e^{-\omega/T} \, d\omega
\nonumber\\&&
-\frac{3}{16} \Big( \langle \bar{q} q \rangle \langle g_s \bar{s} \sigma G s \rangle
+ \langle \bar{s} s \rangle \langle g_s \bar{q} \sigma G q \rangle \Big)\Bigg\},
\label{eq:Kc}
\\
&&f^2 d_M \Sigma\, e^{-\overline{\Lambda} / T}
\nonumber\\&=& \int_{s_<}^{\omega_c}
\Bigg[
\frac{4 g_s^2 \, \omega^7}{105 \, \pi^6}
+ \frac{\langle g^2 GG \rangle \, \omega^3}{48 \, \pi^4}
\nonumber\\&&
- \frac{\omega}{16 \, \pi^2}
\Big(m_q \langle g_s \bar{q} \sigma G q \rangle + m_s \langle g_s \bar{s} \sigma G s \rangle
\nonumber\\&&
- 2m_s \langle g_s \bar{q} \sigma G q \rangle - 2m_q \langle g_s \bar{s} \sigma G s \rangle \Big)
\Bigg] e^{-\omega/T} \, d\omega
\nonumber\\&&
- \frac{1}{24} \Big(
\langle \bar{q} q \rangle \langle g_s \bar{s} \sigma G s \rangle
+ \langle \bar{s} s \rangle \langle g_s \bar{q} \sigma G q \rangle \Big) \, .
\label{eq:Sc}
\end{eqnarray}
The Borel mass dependence of these sum rule equations is shown in Fig.~\ref{fig:KandS}. Both $K$ and $\Sigma$ exhibit mild variation with $T$ in the working region $0.42~\mathrm{GeV} < T < 0.48~\mathrm{GeV}$. From this region, we extract the following numerical results:
\begin{eqnarray}
K &=& -1.11^{+0.18}_{-0.21}~\mathrm{GeV}^2 \, ,
\\
d_M \Sigma &=& 0.27^{+0.04}_{-0.03}~\mathrm{GeV}^2 \, .
\end{eqnarray}

\section{Discussions and Conclusion}
\label{sec:summary}

Based on the results presented in Sec.~\ref{sec:leading} and Sec.~\ref{sec:nexttoleading}, we can determine the mass of the ground-state topped baryon $\Xi^\prime_t$. A crucial consideration in this analysis is the choice of renormalization scheme for the top quark mass. In our previous QCD sum rule studies of charmed and bottom baryons~\cite{Chen:2015kpa,Mao:2015gya,Chen:2017sci,Yang:2020zrh,Yang:2021lce,Wang:2024rai,Luo:2025jpn}, the $\overline{\mathrm{MS}}$ quark masses were commonly adopted:
\begin{eqnarray}
m_c^{\overline{\mathrm{MS}}} &=& 1.2730 _{-0.0046}^{+0.0046}~\mathrm{GeV} \, , 
\\
m_b^{\overline{\mathrm{MS}}} &=& 4.183_{-0.007}^{+0.007}~\mathrm{GeV} \, ,
\end{eqnarray}
with corresponding pole masses:
\begin{eqnarray}
m_c^{\mathrm{pole}} &=& 1.67_{-0.07}^{+0.07}~\mathrm{GeV} \, , 
\\
m_b^{\mathrm{pole}} &=& 4.78_{-0.06}^{+0.06}~\mathrm{GeV} \, ,
\end{eqnarray}
showing moderate differences of several hundred MeV.

For the top quark, however, the discrepancy becomes significantly larger:
\begin{eqnarray}
m_t^{\overline{\mathrm{MS}}} &=& 162.5_{-1.5}^{+2.1}~\mathrm{GeV} \, , 
\\
m_t^{\mathrm{pole}} &=& 172.57_{-0.29}^{+0.29}~\mathrm{GeV} \, ,
\end{eqnarray}
leading to a difference of nearly $10~\mathrm{GeV}$. Given that the residual mass $\overline{\Lambda}$ is only of order $1~\mathrm{GeV}$, such a large mass gap would distort the scale hierarchy and compromise the numerical reliability of the extracted baryon mass. For this reason, we adopt the pole mass of the top quark when evaluating the physical mass of the topped baryon in this work. This choice may also suggest that pole masses are generally more suitable for QCD sum rule analyses—a possibility we intend to explore further in future studies. Additionally, when electroweak corrections are taken into account, the $\overline{\text{MS}}$ mass of the top quark closely approximates its pole mas. Notably, when electroweak corrections are included, the $\overline{\mathrm{MS}}$ top mass approaches the pole mass; see Ref.~\cite{Kataev:2022dua}.

Using Eq.~(\ref{eq:leading}) and Eq.~(\ref{eq:more}), the mass of the ground-state $\Xi^\prime_t$ baryon is calculated as:
\begin{eqnarray}
&& m_{\Xi^\prime_t} 
\\ \nonumber &=& m_t + \overline{\Lambda} + \delta m 
\\ \nonumber &=& 172.57^{+0.29}_{-0.29}~\mathrm{GeV} + 1.38^{+0.14}_{-0.14}~\mathrm{GeV} + 2.8^{+0.6}_{-0.5}~\mathrm{MeV}
\\ \nonumber &=& 173.95^{+0.32}_{-0.32}~\mathrm{GeV} \, .
\end{eqnarray}
This result demonstrates the robustness of the HQET framework, even when extrapolated to the extreme mass regime of the top quark. However, the relatively large uncertainty associated with the top quark mass induces non-negligible sensitivity in the predicted mass of the topped baryon. 

In addition to the $\Xi^\prime_t$ baryon, other ground-state singly topped baryons have been analyzed in a similar manner. Their corresponding sum rule expressions are given in Appendix~\ref{sec:others}, and the numerical results are summarized in Table~\ref{tab:result}. Generally speaking, the masses of ground-state singly topped baryons are calculated to be around $174$~GeV, approximately $1.1$–$1.5$~GeV above the pole mass of the top quark. For comparison, the masses of ground-state singly bottom baryons are measured to be approximately $0.8$–$1.3$~GeV above the pole mass of the bottom quark, and the masses of ground-state singly charmed baryons are measured to be approximately $0.6$–$1.0$~GeV above the pole mass of the charm quark~\cite{pdg}.

\begin{table*}[hbtp]
\begin{center}
\renewcommand{\arraystretch}{2.0}
\caption{Parameters of singly topped baryons calculated using QCD sum rules within the framework of heavy quark effective theory. For the two baryons $\Lambda_t$ and $\Xi_t$, no valid working regions are found; therefore, the convergence criterion given in Eq.~(\ref{eq_convergence}) is used solely to constrain the parameter $T$. The last column lists several representative decay channels.}
\begin{tabular}{ c | c | c | c | c | c | c | c | c}
\hline\hline
\multirow{2}{*}{~~B~~} & $\omega_c$& ~~~~~Working region~~~~~ & ~~~~~~$\overline{\Lambda}$~~~~~~ & ~~Baryon~~ & ~~~~~Mass~~~~~~ & ~Difference~ & ~~~~~~~$f$~~~~~~~ & \multirow{2}{*}{~Decay channels~}
\\                                               & (GeV) & (GeV)      & (GeV)                &                               ($J^P$)       & (GeV)      & (MeV)        & (GeV$^{3}$) & 
\\ \hline\hline
\multirow{2}{*}{$\Sigma_t(\bm{6}_F)$}& \multirow{2}{*}{1.65}& \multirow{2}{*}{$0.40\le T \le0.44$}& \multirow{2}{*}{$1.26 _{-0.13}^{+0.13}$}& $\Sigma_t(1/2^+)$& $173.83_{-0.32}^{+0.32}$& \multirow{2}{*}{$0.64_{-0.07}^{+0.08}$}& $0.07_{-0.02}^{+0.02}$ &\multirow{2}{*}{$\Upsilon \Sigma_c^{(*)}/ \Upsilon \Lambda_c$}
\\ \cline{5-6}\cline{8-8}& & & & $\Sigma_t(3/2^+)$& $
173.83_{-0.32}^{+0.32}$&&$0.04_{-0.01}^{+0.01}$ & 
\\ \hline
 \multirow{2}{*}{$\Xi^\prime_t(\bm{6}_F)$}& \multirow{2}{*}{1.85}& \multirow{2}{*}{$0.42\le T \le 0.48$}& \multirow{2}{*}{$1.38 _{-0.14}^{+0.14}$}& $\Xi^\prime_t(1/2^+)$& $173.95_{-0.32}^{+0.32}$& \multirow{2}{*}{$0.69 _{-0.08}^{+0.10}$}& $0.09_{-0.02}^{+0.03}$ & \multirow{2}{*}{$\Upsilon \Xi_c^{(\prime*)}$} \\ \cline{5-6} \cline{8-8}& & & & $\Xi^\prime_t(3/2^+)$& $173.95_{-0.32}^{+0.32}$& &$0.05_{-0.01}^{+0.02}$ \\ \hline
\multirow{2}{*}{$\Omega_t(\bm{6}_F)$}& \multirow{2}{*}{2.02}& \multirow{2}{*}{$0.43\leq T \leq 0.53$}& \multirow{2}{*}{$1.50 _{-0.15}^{+0.15}$}& $\Omega_t(1/2^+)$& $174.08_{-0.33}^{+0.33}$& \multirow{2}{*}{$0.77_{-0.10}^{+0.12}$}& $0.12_{-0.03}^{+0.04}$ & \multirow{2}{*}{$\Upsilon \Omega_c^{(*)}$}\\ \cline{5-6}\cline{8-8}& & & & $\Omega_t(3/2^+)$& $174.08_{-0.33}^{+0.33}$& &$0.07_{-0.02}^{+0.02}$\\ \hline
 $\Lambda_t(\bm{\bar{3}}_F)$ & 1.41 & $T=0.42, \, {\rm PC}=17\%$ & $1.06 _{-0.15}^{+0.14}$ & $\Lambda_t(1/2^+)$& $173.63_{-0.33}^{+0.32}$& $-$& $0.03 _{-0.01}^{+0.01}$ & $\Upsilon \Sigma_c^{(*)}/ \Upsilon \Lambda_c$\\ \hline
 $\Xi_t(\bm{\bar{3}}_F)$ & 1.62 & $T=0.45,\,{\rm PC}=19\%$ & $1.19_{-0.16}^{+0.14}$ & $\Xi_t(1/2^+)$& $173.76_{-0.33}^{+0.32}$& $-$&$0.04_{-0.01}^{+0.01}$ & $\Upsilon \Xi_c^{(\prime*)}$\\ \hline\hline
\end{tabular}
\label{tab:result}
\end{center}
\end{table*}

In conclusion, this work presents a pioneering application of the QCD sum rule method within the HQET framework to investigate baryons containing top quarks. In parallel, topped mesons are systematically examined in Ref.~\cite{Zhang:2025fdp} using the same theoretical approach. Although both classes of hadrons are unlikely to exist due to the extremely short lifetime of the top quark, their theoretical study enriches our understanding of QCD in the heavy-quark limit. Remarkably, the decay width of the top quark is comparable to those of the shortest-lived hadrons, such as the $\sigma(500)$ and $\kappa(800)$, with estimated widths of $400$--$700~\mathrm{MeV}$ and $500$--$800~\mathrm{MeV}$, respectively~\cite{pdg}, placing it near the boundary where hadronization becomes ineffective. Compared to toponia, singly topped baryons and topped mesons—featuring only one decaying top quark—are expected to have a narrower decay width and thus a relatively longer lifetime. This subtle distinction highlights a transitional domain where weak decay competes with strong binding, offering a unique theoretical setting to probe the onset of hadron formation in the extreme mass limit.

\section*{Acknowledgments}

This project is supported by
the National Natural Science Foundation of China under Grant No.~12075019,
the Jiangsu Provincial Double-Innovation Program under Grant No.~JSSCRC2021488,
the SEU Innovation Capability Enhancement Plan for Doctoral Students,
and
the Fundamental Research Funds for the Central Universities.

\appendix
\begin{widetext}
\section{Other sum rule equations}
\label{sec:others}

In this appendix we present the sum rule equations derived in this study. The expressions associated with the current $J_{\Xi^\prime_t}(x)$ have already been given in Eq.~(\ref{eq:ope}), Eq.~(\ref{eq:Kc}), and Eq.~(\ref{eq:Sc}). The corresponding sum rule equations obtained from the currents $J_{\Sigma_t}(x)$, $J_{\Omega_t}(x)$, $J_{\Lambda_t}(x)$, and $J_{\Xi_t}(x)$ are given below:
\begin{eqnarray}
\label{eq:opesigma}
\Pi_{\Sigma_t}(\omega_c, T) &=& \int_{s_<}^{\omega_c} \Big[\frac{3\omega^{5}}{10\pi^4}
-\frac{3m_{q}^2
\omega^{3}}{2\pi^{4}}-\frac{\langle g^2 GG \rangle
\omega}{64
\pi^4}-\frac{3m_{q}\langle\bar{{q}}{q}\rangle
}{\pi^{2}}\omega\Big]
e^{-\omega/T}{\rm
d}\omega\nonumber\\&&+\langle\bar{{q}}{q}\rangle\langle\bar{{q}}{q}\rangle
- {7m_{q} \langle g_s \bar {q} \sigma G {q}
\rangle
\over 32\pi^2}
+\frac{
\langle\bar{{q}}{q}\rangle\langle g_s\bar{{q}}\sigma
G{q}\rangle  }{8T^2},
\\
f_{ \Sigma_t}^2 K_{ \Sigma_t} e^{-\overline{\Lambda}_{ \Sigma_t} / T}
&=&-\Bigg\{\int^{\omega_{c}}_{s_<}
\Big[\frac{396\omega^7}{7!\pi^4}
-\frac{3m_{q}^2\omega^5}{4\pi^4} -\frac{\langle g^2 GG \rangle \omega^3}{384
\pi^4}-\frac{m_{q}\langle\bar{q}q\rangle\omega^3}{2\pi^2}\nonumber\\&&+\frac{11m_{q}\langle
g_s\bar{q}\sigma G q\rangle\omega}{128\pi^2}\Big]e^{-\omega/T}{\rm
d}\omega-\frac{3\langle\bar{q}q\rangle\langle
g_s\bar{q}\sigma Gq\rangle}{8}\Bigg\}\, ,
\label{eq:KcSigma}
\\
f_{ \Sigma_t}^2 d_M\Sigma_{ \Sigma_t} e^{-\overline{\Lambda}_{ \Sigma_t} / T}
&=&  \int^{\omega_{c}}_{s_<}
\Big[\frac{4g_s^2\omega^7}{105\pi^6}+\frac{\langle g^2 GG \rangle
\omega^3}{48
\pi^4}+\frac{m_{q}\langle
g_s\bar{q}\sigma G q\rangle\omega}{8\pi^2}\Big]e^{-\omega/T}{\rm
d}\omega  -\frac{\langle\bar{q}q\rangle\langle g_{s}\bar{q}\sigma G
q\rangle }{12} \, .
\label{eq:ScSigma}
\\
\label{eq:opeOmega}
\Pi_{\Omega_t}(\omega_c, T) &=& \int_{s_<}^{\omega_c} \Big[\frac{3\omega^{5}}{10\pi^4}
-\frac{3m_{s}^2
\omega^{3}}{2\pi^{4}}-\frac{\langle g^2 GG \rangle
\omega}{64
\pi^4}-\frac{3m_{s}\langle\bar{{s}}{s}\rangle
}{\pi^{2}}\omega\Big]
e^{-\omega/T}{\rm
d}\omega\nonumber\\&&+\langle\bar{{s}}{s}\rangle\langle\bar{{s}}{s}\rangle
- {7m_{s} \langle g_s \bar {s} \sigma G {s}
\rangle
\over 32\pi^2}
+\frac{
\langle\bar{{s}}{s}\rangle\langle g_s\bar{{s}}\sigma
G{s}\rangle  }{8T^2}\,,
\\
f_{\Omega_t}^2 K_{ \Omega_t} e^{-\overline{\Lambda}_{\Omega_t} / T}
&=&-\Bigg\{\int^{\omega_{c}}_{s_<}
\Big[\frac{396\omega^7}{7!\pi^4}
-\frac{3m_{s}^2\omega^5}{4\pi^4} -\frac{\langle g^2 GG \rangle \omega^3}{384
\pi^4}-\frac{m_{s}\langle\bar{s}s\rangle\omega^3}{2\pi^2}\nonumber\\&&+\frac{11m_{s}\langle
g_s\bar{s}\sigma G s\rangle\omega}{128\pi^2}\Big]e^{-\omega/T}{\rm
d}\omega-\frac{3\langle\bar{s}s\rangle\langle
g_s\bar{s}\sigma Gs\rangle}{8}\Bigg\}\, ,
\label{eq:KcOmega}
\\
f_{\Omega_t}^2d_M \Sigma_{ \Omega_t} e^{-\overline{\Lambda}_{\Omega_t} / T}
&=&  \int^{\omega_{c}}_{s_<}
\Big[\frac{4g_s^2\omega^7}{105\pi^6}+\frac{\langle g^2 GG \rangle
\omega^3}{48
\pi^4}+\frac{m_{s}\langle
g_s\bar{s}\sigma G s\rangle\omega}{8\pi^2}\Big]e^{-\omega/T}{\rm
d}\omega  -\frac{\langle\bar{s}s\rangle\langle g_{s}\bar{s}\sigma G
s\rangle }{12} \, .
\label{eq:ScOmega}
\\
\label{eq:opeLambda}
\Pi_{\Lambda_t}(\omega_c, T) &=& \int^{\omega_{c}}_{s_<}\Big[\frac{\omega^{5}}{20\pi^4}-\frac{m_{{q}}^2
{\omega^{3}}}{4\pi^{4}}
+ \frac{\langle g^2 GG \rangle \omega}{128 \pi^4}  -
\frac{m_{q}\langle\bar{{q}}{q}\rangle
 }{2\pi^2
}\omega] e^{-\omega/T}{\rm d}\omega \nonumber\\&& +
\frac{\langle\bar{{q}}{q}\rangle\langle\bar{{q}}{q}\rangle}{6}-11\frac{m_{{q}}\langle
g_{s}\bar{{q}}\sigma G {q}\rangle}{192\pi^2} 
+ \frac{\langle\bar{{q}}{q}\rangle\langle
g_{s}\bar{{q}}\sigma G
{q}\rangle}{48 T^2}\,,
\\
f_{ \Lambda_t}^2 K_{ \Lambda_t} e^{-\overline{\Lambda}_{ \Lambda_t} / T}
&=&-\Bigg\{\int^{\omega_{c}}_{s_<}\Big[\frac{108\omega^7}{7!\pi^4}
-\frac{18m_{q}^2\omega^5}{5!\pi^4}
+\frac{\langle g^2 GG \rangle \omega^3}{128
\pi^4}-\frac{ m_{q}\langle\bar{q}q\rangle\omega^3}{2\pi^2}+\frac{9m_{q}\langle g_s \bar{q}\sigma
Gq\rangle\omega}{32\pi^2}\Big]e^{-\omega/T}{\rm
d}\omega\nonumber\\&&-\frac{\langle\bar{q}q\rangle\langle
  g_{s}\bar{q}\sigma G
 q\rangle}{8}\Bigg\},
\label{eq:KcLambda}
\\
f_{ \Lambda_t}^2 d_M\Sigma_{ \Lambda_t} e^{-\overline{\Lambda}_{ \Lambda_t} / T}
&=& 0.
\label{eq:ScLambda}
\\
\label{eq:opeXi}
\Pi_{\Xi_t}(\omega_c, T) &=& \int^{\omega_{c}}_{s_<}\Big[\frac{\omega^{5}}{20\pi^4}-\frac{(m_{{q}}^2
+m_{{s}}^2-m_{{q}}m_{{s}}){\omega^{3}}}{4\pi^{4}}
+ \frac{\langle g^2 GG \rangle \omega}{128 \pi^4}  \nonumber\\&&+
\frac{m_{{s}}\langle\bar{{s}}{s}\rangle
+m_{{q}}\langle\bar{{q}}{q}\rangle -2m_{{s}}\langle\bar{{q}}{q}\rangle+2m_{{q}}\langle\bar{{s}}{s}\rangle}{4\pi^2
}\omega\Big] e^{-\omega/T}{\rm d}\omega -
\frac{m_{{q}}\langle g_{s}\bar{{s}}\sigma G
{s}\rangle+m_{{s}}\langle g_{s}\bar{{q}}\sigma G
{q}\rangle}{32\pi^2}\nonumber\\&& +\frac{m_{{q}}\langle
g_{s}\bar{{q}}\sigma G {q}\rangle+m_{{s}}\langle
g_{s}\bar{{s}}\sigma G {s}\rangle}{384\pi^2} +
\frac{\langle\bar{{q}}{q}\rangle\langle\bar{{s}}{s}\rangle}{6}
+ \frac{\langle\bar{{q}}{q}\rangle\langle
g_{s}\bar{{s}}\sigma G
{s}\rangle+\langle\bar{{s}}{s}\rangle\langle
g_{s}\bar{{q}}\sigma G {q}\rangle}{96T^2}\,,
\\
f_{\Xi_t}^2 K_{ \Xi_t} e^{-\overline{\Lambda}_{\Xi_t} / T}
&=&-\Bigg\{\int^{\omega_{c}}_{s_<}\Big[\frac{108\omega^7}{7!\pi^4}
-\frac{18\omega^5}{5!\pi^4}(m_{q}^2+m_{s}^2-m_{q} m_{s})
+\frac{\langle g^2 GG \rangle \omega^3}{128
\pi^4}+\frac{\omega^3}{4\pi^2}\Big(m_{q}\langle\bar{q}q\rangle+m_{s}\langle\bar{s}s\rangle\nonumber\\&&-
2m_{s}\langle\bar{q}q\rangle-2m_{q}\langle\bar{s}s\rangle\Big)-\frac{3\omega}{64\pi^2}\Big(m_{q}\langle
g_s\bar{q}\sigma G q\rangle+m_{s}\langle g_s \bar{s}\sigma
G
s\rangle\Big)+\frac{3\omega}{16\pi^2}\Big(m_{q}\langle
g_s\bar{s}\sigma G s\rangle\nonumber\\&&+m_{s}\langle g_s \bar{q}\sigma
G q\rangle\Big)\Big]e^{-\omega/T}{\rm
d}\omega-\frac{1}{16}\Big[ \langle\bar{q}q\rangle\langle
g_{s}\bar{s}\sigma G
s\rangle+\langle\bar{s}s\rangle\langle g_s\bar{q}\sigma
G q\rangle \Big]\Bigg\} ,
\label{eq:KcXi}
\\
f_{\Xi_t}^2 d_M\Sigma_{ \Xi_t} e^{-\overline{\Lambda}_{\Xi_t} / T}
&=& 0.
\label{eq:ScXi}
\end{eqnarray}

\end{widetext}
\bibliographystyle{elsarticle-num}
\bibliography{ref}

\begin{thebibliography}{10}
\expandafter\ifx\csname url\endcsname\relax
  \def\url#1{\texttt{#1}}\fi
\expandafter\ifx\csname urlprefix\endcsname\relax\def\urlprefix{URL }\fi
\expandafter\ifx\csname href\endcsname\relax
  \def\href#1#2{#2} \def\path#1{#1}\fi

\bibitem{CMS:2025kzt}
A.~Hayrapetyan, et~al., {Observation of a pseudoscalar excess at the top quark pair production threshold} (3 2025).
\newblock \href {http://arxiv.org/abs/2503.22382} {\path{arXiv:2503.22382}}.

\bibitem{Zhang:2025fdp}
S.-W. Zhang, X.~Luo, H.-M. Yang, H.-X. Chen, {QCD sum rule study of topped mesons within heavy quark effective theory} (8 2025).
\newblock \href {http://arxiv.org/abs/2508.03422} {\path{arXiv:2508.03422}}.

\bibitem{Fadin:1987wz}
V.~S. Fadin, V.~A. Khoze, {Threshold Behavior of Heavy Top Production in $e^+ e^-$ Collisions}, JETP Lett. 46 (1987) 525--529.

\bibitem{Kuhn:1987ty}
J.~H. Kuhn, P.~M. Zerwas, {The Toponium Scenario}, Phys. Rept. 167 (1988) 321.
\newblock \href {https://doi.org/10.1016/0370-1573(88)90075-0} {\path{doi:10.1016/0370-1573(88)90075-0}}.

\bibitem{Barger:1987xg}
V.~D. Barger, E.~W.~N. Glover, K.~Hikasa, W.-Y. Keung, M.~G. Olsson, C.~J. Suchyta, III, X.~R. Tata, {Superheavy Quarkonium Production and Decays: A New Higgs Signal}, Phys. Rev. D 35 (1987) 3366, [Erratum: Phys.Rev.D 38, 1632 (1988)].
\newblock \href {https://doi.org/10.1103/PhysRevD.35.3366} {\path{doi:10.1103/PhysRevD.35.3366}}.

\bibitem{Fadin:1990wx}
V.~S. Fadin, V.~A. Khoze, T.~Sjostrand, {On the Threshold Behavior of Heavy Top Production}, Z. Phys. C 48 (1990) 613--622.
\newblock \href {https://doi.org/10.1007/BF01614696} {\path{doi:10.1007/BF01614696}}.

\bibitem{Strassler:1990nw}
M.~J. Strassler, M.~E. Peskin, {The Heavy top quark threshold: QCD and the Higgs}, Phys. Rev. D 43 (1991) 1500--1514.
\newblock \href {https://doi.org/10.1103/PhysRevD.43.1500} {\path{doi:10.1103/PhysRevD.43.1500}}.

\bibitem{Sumino:1997ve}
Y.~Sumino, {Top quark pair production and decay near threshold in $e^+ e^-$ collisions}, Acta Phys. Polon. B 28 (1997) 2461--2478.
\newblock \href {http://arxiv.org/abs/hep-ph/9711233} {\path{arXiv:hep-ph/9711233}}.

\bibitem{Hoang:2000yr}
A.~H. Hoang, et~al., {Top-Antitop Pair Production close to Threshold: Synopsis of recent NNLO Results}, Eur. Phys. J. direct 2~(1) (2000) 3.
\newblock \href {http://arxiv.org/abs/hep-ph/0001286} {\path{arXiv:hep-ph/0001286}}, \href {https://doi.org/10.1007/s1010500c0003} {\path{doi:10.1007/s1010500c0003}}.

\bibitem{Penin:2005eu}
A.~A. Penin, V.~A. Smirnov, M.~Steinhauser, {Heavy quarkonium spectrum and production/annihilation rates to order $\beta_0^3\alpha_s^3$}, Nucl. Phys. B 716 (2005) 303--318.
\newblock \href {http://arxiv.org/abs/hep-ph/0501042} {\path{arXiv:hep-ph/0501042}}, \href {https://doi.org/10.1016/j.nuclphysb.2005.03.028} {\path{doi:10.1016/j.nuclphysb.2005.03.028}}.

\bibitem{Kiyo:2008bv}
Y.~Kiyo, J.~H. Kuhn, S.~Moch, M.~Steinhauser, P.~Uwer, {Top-quark pair production near threshold at LHC}, Eur. Phys. J. C 60 (2009) 375--386.
\newblock \href {http://arxiv.org/abs/0812.0919} {\path{arXiv:0812.0919}}, \href {https://doi.org/10.1140/epjc/s10052-009-0892-7} {\path{doi:10.1140/epjc/s10052-009-0892-7}}.

\bibitem{Hagiwara:2008df}
K.~Hagiwara, Y.~Sumino, H.~Yokoya, {Bound-state Effects on Top Quark Production at Hadron Colliders}, Phys. Lett. B 666 (2008) 71--76.
\newblock \href {http://arxiv.org/abs/0804.1014} {\path{arXiv:0804.1014}}, \href {https://doi.org/10.1016/j.physletb.2008.07.006} {\path{doi:10.1016/j.physletb.2008.07.006}}.

\bibitem{Sumino:2010bv}
Y.~Sumino, H.~Yokoya, {Bound-state effects on kinematical distributions of top quarks at hadron colliders}, JHEP 09 (2010) 034, [Erratum: JHEP 06, 037 (2016)].
\newblock \href {http://arxiv.org/abs/1007.0075} {\path{arXiv:1007.0075}}, \href {https://doi.org/10.1007/JHEP09(2010)034} {\path{doi:10.1007/JHEP09(2010)034}}.

\bibitem{Beneke:2015kwa}
M.~Beneke, Y.~Kiyo, P.~Marquard, A.~Penin, J.~Piclum, M.~Steinhauser, {Next-to-Next-to-Next-to-Leading Order QCD Prediction for the Top Antitop $S$-Wave Pair Production Cross Section Near Threshold in $e^+e^-$ Annihilation}, Phys. Rev. Lett. 115~(19) (2015) 192001.
\newblock \href {http://arxiv.org/abs/1506.06864} {\path{arXiv:1506.06864}}, \href {https://doi.org/10.1103/PhysRevLett.115.192001} {\path{doi:10.1103/PhysRevLett.115.192001}}.

\bibitem{Fuks:2021xje}
B.~Fuks, K.~Hagiwara, K.~Ma, Y.-J. Zheng, {Signatures of toponium formation in LHC run 2 data}, Phys. Rev. D 104~(3) (2021) 034023.
\newblock \href {http://arxiv.org/abs/2102.11281} {\path{arXiv:2102.11281}}, \href {https://doi.org/10.1103/PhysRevD.104.034023} {\path{doi:10.1103/PhysRevD.104.034023}}.

\bibitem{Akbar:2024brg}
N.~Akbar, I.~Asghar, Z.~Ahmad, {Properties of Toponium Mesons with Non-relativistic QCD Potential Model} (11 2024).
\newblock \href {http://arxiv.org/abs/2411.08548} {\path{arXiv:2411.08548}}.

\bibitem{Wang:2024hzd}
G.-L. Wang, T.-F. Feng, Y.-Q. Wang, {Mass spectra and wave functions of toponia}, Phys. Rev. D 111~(9) (2025) 096016.
\newblock \href {http://arxiv.org/abs/2411.17955} {\path{arXiv:2411.17955}}, \href {https://doi.org/10.1103/PhysRevD.111.096016} {\path{doi:10.1103/PhysRevD.111.096016}}.

\bibitem{Jiang:2024fyw}
S.-J. Jiang, B.-Q. Li, G.-Z. Xu, K.-Y. Liu, {Study on Toponium: Spectrum and Associated Processes} (12 2024).
\newblock \href {http://arxiv.org/abs/2412.18527} {\path{arXiv:2412.18527}}.

\bibitem{Fuks:2024yjj}
B.~Fuks, K.~Hagiwara, K.~Ma, Y.-J. Zheng, {Simulating toponium formation signals at the LHC}, Eur. Phys. J. C 85~(2) (2025) 157.
\newblock \href {http://arxiv.org/abs/2411.18962} {\path{arXiv:2411.18962}}, \href {https://doi.org/10.1140/epjc/s10052-025-13853-3} {\path{doi:10.1140/epjc/s10052-025-13853-3}}.

\bibitem{ATLAS:2023fsd}
G.~Aad, et~al., {Observation of quantum entanglement with top quarks at the ATLAS detector}, Nature 633~(8030) (2024) 542--547.
\newblock \href {http://arxiv.org/abs/2311.07288} {\path{arXiv:2311.07288}}, \href {https://doi.org/10.1038/s41586-024-07824-z} {\path{doi:10.1038/s41586-024-07824-z}}.

\bibitem{CMS:2024pts}
A.~Hayrapetyan, et~al., {Observation of quantum entanglement in top quark pair production in proton{\textendash}proton collisions at $\sqrt{s} = 13$ TeV}, Rept. Prog. Phys. 87~(11) (2024) 117801.
\newblock \href {http://arxiv.org/abs/2406.03976} {\path{arXiv:2406.03976}}, \href {https://doi.org/10.1088/1361-6633/ad7e4d} {\path{doi:10.1088/1361-6633/ad7e4d}}.

\bibitem{Jafari:2025rmm}
A.~Jafari, {TOP2024: an overview of experimental results}, in: {17th International Workshop on Top Quark Physics}, 2025.
\newblock \href {http://arxiv.org/abs/2501.16231} {\path{arXiv:2501.16231}}.

\bibitem{Aguilar-Saavedra:2024mnm}
J.~A. Aguilar-Saavedra, {Toponium hunter{\textquoteright}s guide}, Phys. Rev. D 110~(5) (2024) 054032.
\newblock \href {http://arxiv.org/abs/2407.20330} {\path{arXiv:2407.20330}}, \href {https://doi.org/10.1103/PhysRevD.110.054032} {\path{doi:10.1103/PhysRevD.110.054032}}.

\bibitem{Llanes-Estrada:2024phk}
F.~J. Llanes-Estrada, {Ensuring that toponium is glued, not nailed}, Phys. Lett. B 866 (2025) 139510.
\newblock \href {http://arxiv.org/abs/2411.19180} {\path{arXiv:2411.19180}}, \href {https://doi.org/10.1016/j.physletb.2025.139510} {\path{doi:10.1016/j.physletb.2025.139510}}.

\bibitem{Nason:2025hix}
P.~Nason, E.~Re, L.~Rottoli, {Spin Correlations in $t{\bar t}$ Production and Decay at the LHC in QCD Perturbation Theory} (4 2025).
\newblock \href {http://arxiv.org/abs/2505.00096} {\path{arXiv:2505.00096}}.

\bibitem{Ellis:2025nkm}
J.~Ellis, {Personal Memories of 50 Years of Quarkonia}, in: {50 Years Discovery of the $J/{\ensuremath{\psi}}$ Particle (2024)}, 2025.
\newblock \href {http://arxiv.org/abs/2506.10643} {\path{arXiv:2506.10643}}.

\bibitem{Fu:2025yft}
J.-H. Fu, Y.-J. Li, H.-M. Yang, Y.-B. Li, Y.-J. Zhang, C.-P. Shen, {Toponium: The smallest bound state and simplest hadron in quantum mechanics}, Phys. Rev. D 111~(11) (2025) 114020.
\newblock \href {http://arxiv.org/abs/2412.11254} {\path{arXiv:2412.11254}}, \href {https://doi.org/10.1103/fqc9-k315} {\path{doi:10.1103/fqc9-k315}}.

\bibitem{Fu:2025zxb}
J.-H. Fu, Y.-J. Zhang, G.-Z. Xu, K.-Y. Liu, {Toponium: Implementation of a toponium model in FeynRules} (4 2025).
\newblock \href {http://arxiv.org/abs/2504.12634} {\path{arXiv:2504.12634}}.

\bibitem{Bai:2025buy}
Y.~Bai, T.-K. Chen, Y.~Yang, {Toponia at the HL-LHC, CEPC, and FCC-ee} (6 2025).
\newblock \href {http://arxiv.org/abs/2506.14552} {\path{arXiv:2506.14552}}.

\bibitem{Chen:2021erj}
H.-X. Chen, {Hadronic molecules in $B$ decays}, Phys. Rev. D 105~(9) (2022) 094003.
\newblock \href {http://arxiv.org/abs/2103.08586} {\path{arXiv:2103.08586}}, \href {https://doi.org/10.1103/PhysRevD.105.094003} {\path{doi:10.1103/PhysRevD.105.094003}}.

\bibitem{Maltoni:2024csn}
F.~Maltoni, C.~Severi, S.~Tentori, E.~Vryonidou, {Quantum tops at circular lepton colliders}, JHEP 09 (2024) 001.
\newblock \href {http://arxiv.org/abs/2404.08049} {\path{arXiv:2404.08049}}, \href {https://doi.org/10.1007/JHEP09(2024)001} {\path{doi:10.1007/JHEP09(2024)001}}.

\bibitem{pdg}
S.~Navas, et~al., {Review of particle physics}, Phys. Rev. D 110~(3) (2024) 030001.
\newblock \href {https://doi.org/10.1103/PhysRevD.110.030001} {\path{doi:10.1103/PhysRevD.110.030001}}.

\bibitem{Neubert:1993mb}
M.~Neubert, {Heavy quark symmetry}, Phys. Rept. 245 (1994) 259--396.
\newblock \href {http://arxiv.org/abs/hep-ph/9306320} {\path{arXiv:hep-ph/9306320}}, \href {https://doi.org/10.1016/0370-1573(94)90091-4} {\path{doi:10.1016/0370-1573(94)90091-4}}.

\bibitem{Manohar:2000dt}
A.~V. Manohar, M.~B. Wise, {Heavy quark physics}, Vol.~10, 2000.
\newblock \href {https://doi.org/10.1017/9781009402125} {\path{doi:10.1017/9781009402125}}.

\bibitem{Dai:1993kt}
Y.-B. Dai, C.-S. Huang, H.-Y. Jin, {Bethe-Salpeter wave functions and transition amplitudes for heavy mesons}, Z. Phys. C 60 (1993) 527--534.
\newblock \href {https://doi.org/10.1007/BF01560051} {\path{doi:10.1007/BF01560051}}.

\bibitem{Dai:1996yw}
Y.-B. Dai, C.-S. Huang, M.-Q. Huang, C.~Liu, {QCD sum rules for masses of excited heavy mesons}, Phys. Lett. B 390 (1997) 350--358.
\newblock \href {http://arxiv.org/abs/hep-ph/9609436} {\path{arXiv:hep-ph/9609436}}, \href {https://doi.org/10.1016/S0370-2693(96)01412-8} {\path{doi:10.1016/S0370-2693(96)01412-8}}.

\bibitem{Liu:2007fg}
X.~Liu, H.-X. Chen, Y.-R. Liu, A.~Hosaka, S.-L. Zhu, {Bottom baryons}, Phys. Rev. D 77 (2008) 014031.
\newblock \href {http://arxiv.org/abs/0710.0123} {\path{arXiv:0710.0123}}, \href {https://doi.org/10.1103/PhysRevD.77.014031} {\path{doi:10.1103/PhysRevD.77.014031}}.

\bibitem{Yang:2025hzc}
H.-M. Yang, X.~Luo, H.-X. Chen, W.~Chen, {Investigating charmed hybrid baryons via QCD sum rules} (5 2025).
\newblock \href {http://arxiv.org/abs/2505.18717} {\path{arXiv:2505.18717}}.

\bibitem{Dai:1996qx}
Y.-B. Dai, C.-S. Huang, M.-Q. Huang, {${\mathcal O} (1 / m_Q)$ order corrections to masses of excited heavy mesons from QCD sum rules}, Phys. Rev. D 55 (1997) 5719--5726.
\newblock \href {http://arxiv.org/abs/hep-ph/9702384} {\path{arXiv:hep-ph/9702384}}, \href {https://doi.org/10.1103/PhysRevD.55.5719} {\path{doi:10.1103/PhysRevD.55.5719}}.

\bibitem{Dai:2003yg}
Y.-B. Dai, C.-S. Huang, C.~Liu, S.-L. Zhu, {Understanding the $D^+_{sJ}(2317)$ and $D^+_{sJ}(2460)$ with sum rules in HQET}, Phys. Rev. D 68 (2003) 114011.
\newblock \href {http://arxiv.org/abs/hep-ph/0306274} {\path{arXiv:hep-ph/0306274}}, \href {https://doi.org/10.1103/PhysRevD.68.114011} {\path{doi:10.1103/PhysRevD.68.114011}}.

\bibitem{Yang:1993bp}
K.-C. Yang, W.~Y.~P. Hwang, E.~M. Henley, L.~S. Kisslinger, {QCD sum rules and neutron-proton mass difference}, Phys. Rev. D 47 (1993) 3001--3012.
\newblock \href {https://doi.org/10.1103/PhysRevD.47.3001} {\path{doi:10.1103/PhysRevD.47.3001}}.

\bibitem{Narison:2011xe}
S.~Narison, {Gluon condensates and precise $\overline{m}_{c,b}$ from QCD-moments and their ratios to order $\alpha_s^3$ and $\langle G^4 \rangle$}, Phys. Lett. B 706 (2012) 412--422.
\newblock \href {http://arxiv.org/abs/1105.2922} {\path{arXiv:1105.2922}}, \href {https://doi.org/10.1016/j.physletb.2011.11.058} {\path{doi:10.1016/j.physletb.2011.11.058}}.

\bibitem{Narison:2018dcr}
S.~Narison, {QCD parameter correlations from heavy quarkonia}, Int. J. Mod. Phys. A 33~(10) (2018) 1850045, [Addendum: Int.J.Mod.Phys.A 33, 1892004 (2018)].
\newblock \href {http://arxiv.org/abs/1801.00592} {\path{arXiv:1801.00592}}, \href {https://doi.org/10.1142/S0217751X18500458} {\path{doi:10.1142/S0217751X18500458}}.

\bibitem{Gimenez:2005nt}
V.~Gimenez, V.~Lubicz, F.~Mescia, V.~Porretti, J.~Reyes, {Operator product expansion and quark condensate from lattice QCD in coordinate space}, Eur. Phys. J. C 41 (2005) 535--544.
\newblock \href {http://arxiv.org/abs/hep-lat/0503001} {\path{arXiv:hep-lat/0503001}}, \href {https://doi.org/10.1140/epjc/s2005-02250-9} {\path{doi:10.1140/epjc/s2005-02250-9}}.

\bibitem{Jamin:2002ev}
M.~Jamin, {Flavor-symmetry breaking of the quark condensate and chiral corrections to the Gell-Mann-Oakes-Renner relation}, Phys. Lett. B 538 (2002) 71--76.
\newblock \href {http://arxiv.org/abs/hep-ph/0201174} {\path{arXiv:hep-ph/0201174}}, \href {https://doi.org/10.1016/S0370-2693(02)01951-2} {\path{doi:10.1016/S0370-2693(02)01951-2}}.

\bibitem{Ioffe:2002be}
B.~L. Ioffe, K.~N. Zyablyuk, {Gluon condensate in charmonium sum rules with three-loop corrections}, Eur. Phys. J. C 27 (2003) 229--241.
\newblock \href {http://arxiv.org/abs/hep-ph/0207183} {\path{arXiv:hep-ph/0207183}}, \href {https://doi.org/10.1140/epjc/s2002-01099-8} {\path{doi:10.1140/epjc/s2002-01099-8}}.

\bibitem{Ovchinnikov:1988gk}
A.~A. Ovchinnikov, A.~A. Pivovarov, {QCD sum rule calculation of the quark gluon condensate}, Sov. J. Nucl. Phys. 48 (1988) 721--723.

\bibitem{Colangelo:2000dp}
P.~Colangelo, A.~Khodjamirian, {QCD sum rules, a modern perspective} (2000) 1495--1576\href {http://arxiv.org/abs/hep-ph/0010175} {\path{arXiv:hep-ph/0010175}}, \href {https://doi.org/10.1142/9789812810458_0033} {\path{doi:10.1142/9789812810458_0033}}.

\bibitem{Chen:2015kpa}
H.-X. Chen, W.~Chen, Q.~Mao, A.~Hosaka, X.~Liu, S.-L. Zhu, {$P$-wave charmed baryons from QCD sum rules}, Phys. Rev. D 91~(5) (2015) 054034.
\newblock \href {http://arxiv.org/abs/1502.01103} {\path{arXiv:1502.01103}}, \href {https://doi.org/10.1103/PhysRevD.91.054034} {\path{doi:10.1103/PhysRevD.91.054034}}.

\bibitem{Mao:2015gya}
Q.~Mao, H.-X. Chen, W.~Chen, A.~Hosaka, X.~Liu, S.-L. Zhu, {QCD sum rule calculation for $P$-wave bottom baryons}, Phys. Rev. D 92~(11) (2015) 114007.
\newblock \href {http://arxiv.org/abs/1510.05267} {\path{arXiv:1510.05267}}, \href {https://doi.org/10.1103/PhysRevD.92.114007} {\path{doi:10.1103/PhysRevD.92.114007}}.

\bibitem{Chen:2017sci}
H.-X. Chen, Q.~Mao, W.~Chen, A.~Hosaka, X.~Liu, S.-L. Zhu, {Decay properties of $P$-wave charmed baryons from light-cone QCD sum rules}, Phys. Rev. D 95~(9) (2017) 094008.
\newblock \href {http://arxiv.org/abs/1703.07703} {\path{arXiv:1703.07703}}, \href {https://doi.org/10.1103/PhysRevD.95.094008} {\path{doi:10.1103/PhysRevD.95.094008}}.

\bibitem{Yang:2020zrh}
H.-M. Yang, H.-X. Chen, {$P$-wave bottom baryons of the $SU(3)$ flavor $\mathbf{6}_F$}, Phys. Rev. D 101~(11) (2020) 114013, [Erratum: Phys.Rev.D 102, 079901 (2020)].
\newblock \href {http://arxiv.org/abs/2003.07488} {\path{arXiv:2003.07488}}, \href {https://doi.org/10.1103/PhysRevD.101.114013} {\path{doi:10.1103/PhysRevD.101.114013}}.

\bibitem{Yang:2021lce}
H.-M. Yang, H.-X. Chen, {$P$-wave charmed baryons of the $SU(3)$ flavor $\mathbf{6}_F$}, Phys. Rev. D 104~(3) (2021) 034037.
\newblock \href {http://arxiv.org/abs/2106.15488} {\path{arXiv:2106.15488}}, \href {https://doi.org/10.1103/PhysRevD.104.034037} {\path{doi:10.1103/PhysRevD.104.034037}}.

\bibitem{Wang:2024rai}
Y.-J. Wang, X.~Luo, H.-X. Chen, E.-L. Cui, W.-H. Tan, Z.-Y. Zhou, {Strong decay properties of $P$-wave single bottom baryons of the $SU(3)$ flavor antitriplet $\bm{\bar{3}}_F$}, Phys. Rev. D 111~(7) (2025) 076003.
\newblock \href {http://arxiv.org/abs/2412.19846} {\path{arXiv:2412.19846}}, \href {https://doi.org/10.1103/PhysRevD.111.076003} {\path{doi:10.1103/PhysRevD.111.076003}}.

\bibitem{Luo:2025jpn}
X.~Luo, Y.-J. Wang, H.-X. Chen, {$P$-wave single charmed baryons of the $SU(3)$ flavor $\bm{\bar{3}}_F$}, Phys. Rev. D 111~(9) (2025) 094039.
\newblock \href {http://arxiv.org/abs/2504.11219} {\path{arXiv:2504.11219}}, \href {https://doi.org/10.1103/p48x-mbnm} {\path{doi:10.1103/p48x-mbnm}}.

\bibitem{Kataev:2022dua}
A.~L. Kataev, V.~S. Molokoedov, {Notes on Interplay between the QCD and EW Perturbative Corrections to the Pole-Running-to-Top-Quark Mass Ratio}, JETP Lett. 115~(12) (2022) 704--712.
\newblock \href {http://arxiv.org/abs/2201.12073} {\path{arXiv:2201.12073}}, \href {https://doi.org/10.1134/S0021364022600902} {\path{doi:10.1134/S0021364022600902}}.

\end{thebibliography}

\end{document}